%% file: BLASTTNG_SPIE2016_arxiv.tex
\documentclass[]{spie}  

\usepackage{amsmath,amsfonts,amssymb}
\usepackage{graphicx}
\usepackage[colorlinks=true, allcolors=blue]{hyperref}
\usepackage{aas_macros}

\newcommand{\mum}{\,$\mu$m}
\newcommand{\arcsec}{$^{\prime\prime}$}
\newcommand{\arcmin}{$^{\prime}$}

\title{Instrumental performance and results from testing of the BLAST-TNG receiver, submillimeter optics, and MKID arrays}

\author[a]{Nicholas Galitzki}
\author[b]{Peter Ade}
\author[a]{Francesco E. Angil\`e}
\author[c]{Peter Ashton}
\author[d]{Jason Austermann}
\author[a]{Tashalee Billings}
\author[e]{George Che}
\author[f]{Hsiao-Mei Cho}
\author[e]{Kristina Davis}
\author[a]{Mark Devlin}
\author[a]{Simon Dicker}
\author[a]{Bradley J. Dober}
\author[c]{Laura M. Fissel}
\author[g]{Yasuo Fukui}
\author[d]{Jiansong Gao}
\author[e]{Samuel Gordon}
\author[e]{Christopher E. Groppi}
\author[a]{Seth Hillbrand}
\author[d]{Gene C. Hilton}
\author[d]{Johannes Hubmayr}
\author[h]{Kent D. Irwin}
\author[a]{Jeffrey Klein}
\author[f]{Dale Li}
\author[i]{Zhi-Yun Li}
\author[a]{Nathan P. Lourie}
\author[a]{Ian Lowe}
\author[e]{Hamdi Mani}
\author[j]{Peter G. Martin}
\author[e]{Philip Mauskopf}
\author[d]{Christopher McKenney}
\author[a]{Federico Nati}
\author[c]{Giles Novak}
\author[b]{Enzo Pascale}
\author[b]{Giampaolo Pisano}
\author[c]{Fabio P. Santos}
\author[k]{Douglas Scott}
\author[e]{Adrian Sinclair}
\author[l]{Juan D. Soler}
\author[b]{Carole Tucker}
\author[e]{Matthew Underhill}
\author[d]{Michael Vissers}
\author[c]{Paul Williams}

\affil[a]{University of Pennsylvania, 209 S 33rd St, Philadelphia, PA, USA}
\affil[b]{University of Cardiff, The Parade, Cardiff, United Kingdom}
\affil[c]{Center for Interdisciplinary Exploration and Research in Astrophysics (CIERA) and Department of Physics \& Astronomy, Northwestern University, 2145 Sheridan Road, Evanston, IL 60208, USA
}
\affil[d]{National Institute of Standards and Technology, 325 Broadway, Boulder, CO, USA}
\affil[e]{Arizona State University, PO Box 871404, Tempe, AZ, USA}
\affil[f]{SLAC National Accelerator Laboratory, 2575 Sand Hill Road, Menlo Park, CA, USA}
\affil[g]{Nagoya University, Furocho, Chikusa Ward, Nagoya, Aichi Prefecture, Japan}
\affil[h]{Stanford University, 382 Via Pueblo Mall, Stanford, CA, USA}
\affil[i]{University of Virginia, 530 McCormick Road, Charlottesville, VA, USA}
\affil[j]{University of Toronto, 60 St. George Street, Toronto, ON, Canada}
\affil[l]{University of British Columbia, 6224 Agricultural Road, Vancouver, BC, Canada}
\affil[k]{Laboratoire AIM, Paris-Saclay, CEA/IRFU/SAp - CNRS - Universit\'{e} Paris Diderot, 91191, Gif-sur-Yvette Cedex, France}

\authorinfo{E-mail: galitzki@sas.upenn.edu, Telephone: 1 215 573 7558}

\begin{document} 
\maketitle

\begin{abstract}
Polarized thermal emission from interstellar dust grains can be used to map magnetic fields in star forming molecular clouds and the diffuse interstellar medium (ISM). The Balloon-borne Large Aperture Submillimeter Telescope for Polarimetry (BLASTPol) flew from Antarctica in 2010 and 2012 and produced degree-scale polarization maps of several nearby molecular clouds with arcminute resolution. The success of BLASTPol has motivated a next-generation instrument, BLAST-TNG, which will use more than 3000 linear polarization-sensitive microwave kinetic inductance detectors (MKIDs) combined with a 2.5\,m diameter carbon fiber primary mirror to make diffraction-limited observations at 250, 350, and 500\mum. With 16 times the mapping speed of BLASTPol, sub-arcminute resolution, and a longer flight time, BLAST-TNG will be able to examine nearby molecular clouds and the diffuse galactic dust polarization spectrum in unprecedented detail. The 250\mum\ detector array has been integrated into the new cryogenic receiver, and is undergoing testing to establish the optical and polarization characteristics of the instrument. BLAST-TNG will demonstrate the effectiveness of kilo-pixel MKID arrays for applications in submillimeter astronomy. BLAST-TNG is scheduled to fly from Antarctica in December 2017 for 28 days and will be the first balloon-borne telescope to offer a quarter of the flight for ``shared risk'' observing by the community.
\end{abstract}


\section{INTRODUCTION}
\label{sec:intro}  

\input{blast-tng_model}

The Balloon-borne Large Aperture Submillimeter Telescope - The Next Generation (BLAST-TNG)\cite{Galitzki2014b} consists of a 2.5\,m Cassegrain telescope with three microwave kinetic inductance detector (MKID) arrays operating over 30\% bandwidths centered on 250, 350, and 500\mum, which have 1836, 950, 460 detectors with diffraction limited resolution of 25\arcsec, 35\arcsec, and 50\arcsec, respectively. It will be flown on a stratospheric balloon as part of NASA's long duration balloon program. An overview of the instrument design is shown in Figure \ref{fig:model}.

The primary science goal of BLAST-TNG is to probe polarized thermal emission from dust in Galactic star forming regions in order to detect the local magnetic field orientation as projected on to the plane of the sky. The Antarctic flights in 2010\cite{Pascale2012} and 2012\cite{Galitzki2014} of the previous instrument, BLASTPol, produced maps of magnetic field morphology in star forming regions with unprecedented levels of detail\cite{Matthews2014, Gandilo2016, Fissel2016, Santos2016, Soler2016, Galitzki2016, Shariff2016}. The demonstrated scientific potential of submillimeter polarimetry observations prompted the design and construction of BLAST-TNG, which will have 16 times the BLASTPol mapping speed as well as better resolution.

The increased collecting area, sensitivity, and resolution will allow for the mapping of more clouds than the 2012 flight and enable observations of diffuse galactic dust emission. BLAST-TNG will allow us to investigate the role magnetic fields play in the formation and evolution of molecular clouds and their associated sub-structures including filaments, cores, and protostars as well as furthering our understanding of dust grain emission and alignment mechanisms. Additionally, BLAST-TNG provides a unique platform for observations which link the large-scale {\it Planck}\cite{PlanckI2015} polarimetry maps to the small scale, but high resolution, polarimetry maps made by interferometers such as ALMA\cite{ALMA2016}.

\section{Magnetic Fields in Star-Forming Regions}
\label{sec:magfield}  

In order to create a complete theory of star formation we must understand the processes that regulate the star formation rate in molecular clouds. Recent progress includes using observations of dust emission and extinction, which shows how core mass distribution correlates with observed stellar mass distribution\cite{Nutter2007}. {\it Herschel} data has yielded many important results relevant to the star formation process such as observations that filamentary structures show a tendency for the densest filaments to be perpendicular to the local magnetic field, while faint, low density sub-filaments tend to be parallel to the local magnetic field\cite{Palmeirim2013}. Additionally, {\it Planck}'s all-sky maps of polarized dust emission have greatly increased our understanding of the role magnetic fields play in our galaxy. From these maps, strong evidence has been found that magnetic field orientation is correlated with molecular cloud structure and the degree of this alignment depends on the magnetic field strength and turbulent properties of the cloud being observed\cite{PlanckXXXV2016}.
 
However, there are many questions concerning the star-formation process and the evolution of cloud structure that remain to be addressed\cite{McKee2007}. Examples include whether the lifetimes of molecular clouds and their internal structures are equal\cite{Vazquez2006} to or larger\cite{Netterfield2009,Blitz2007,Goldsmith2008} than the turbulent crossing time. To have lifetimes longer than crossing times would require a supporting mechanism to counteract gravity. Magnetic fields could provide such support and numerical simulations have shown magnetic fields in clouds can drastically alter star formation efficiencies and the lifetimes of molecular clouds\cite{Li2010,Hennebelle2011}.  However, knowledge of magnetic fields and their interaction with molecular cloud structure is still fairly limited. Zeeman splitting observations have produced measurements of the field strength along the line of sight, but are limited to bright regions, and optical extinction polarization observations have produced a small number of magnetic field pseudo-vectors,  but only in areas of low extinction\cite{Crutcher2010,Falgarone2008}. 

The most promising method for detecting magnetic fields over large ranges of dust column density is with far-IR and submillimeter polarimetry\cite{Hildebrand2000,WardThompson2000,WardThompson2009}. Spinning dust grains preferentially align with short axes parallel to the local magnetic field through the process of radiative alignment torques\cite{Andersson2015}. The grains emit modified blackbody radiation that peaks in the far-IR/submillimeter and is polarized orthogonally to the local magnetic field. BLAST-TNG will have the unique capability to create degree-scale polarization maps of molecular clouds with sub-arcminute resolution with a mapping speed that will allow it to cover multiple targets during each flight.  BLAST-TNG data will build on the results of BLASTPol, enabling additional and more detailed comparisons between polarization maps and numerical simulations\cite{WardThompson2000}.

BLAST-TNG observations will target the following three key questions in star formation:\emph{
}i) \emph{Is core morphology and evolution determined by large-scale magnetic fields?} ii) \emph{Does filamentary structure have
a magnetic origin? iii) What is the field strength, and how does it vary from cloud to cloud?} as further discussed in Ref. \citenum{Fissel2010}.

\section{Instrument} 
\label{sec:instrument} 

\input{fgr_optic_v1}

BLAST-TNG will continue the legacy of BLASTPol with the construction of a new instrument that incorporates many successful elements from previous ballooning experiments. The optical layout of BLAST-TNG is shown in Figure \ref{fig:optics} with the optical component parameters listed in Table \ref{tab:optica}. A 2.5\,m diameter carbon fiber primary mirror\footnote{Vanguard Space Technologies: 9431 Dowdy Drive, San Diego, CA 92126} provides diffraction-limited observations. The telescope uses a Cassegrain configuration with the primary mirror illuminating a 56\,cm diameter aluminum secondary mirror. The secondary mirror is actuated to allow for refocusing during the flight to adjust to the differential thermal contraction of the support structure. The beam is then re-imaged by the 4\,K cold optics which use a modified Offner relay configuration. The light is split by two dichroic filters\cite{Ade2006} into the science bands at 250, 350, and 500\mum\ to allow for simultaneous observations of the same patch of sky. The cold optics have been installed in the flight cryostat and will be characterized concurrently with the detector performance. The primary and secondary mirrors and supporting structure are on schedule for a July 2016 delivery.

Achieving a circular, 22\arcmin\ diameter, field of view (FOV) requires re-imaging optics significantly larger than those of BLASTPol. The size constraints necessitated the construction of a new cryostat which was also designed to have a longer hold time of $\sim$28 days, versus the previous 13 day hold time of BLASTPol. The BLAST-TNG cryostat has a 250 liter liquid helium bath to provide cooling to 4\,K and utilizes two vapor-cooled shields (VCS) to provide additional thermal isolation. The arrays are kept at 270\,mK by a closed-cycle \textsuperscript{3}He refrigerator with a 1\,K intercept stage provided by an open cycle \textsuperscript{4}He pumped pot that is fed off the main Helium tank and vents to atmosphere during flight. The detector arrays and cold optics are contained within an Amuneal\footnote{Amuneal Manufacturing Corp., 4737 Darrah Street, Philadelphia, PA 19124} enclosure which provides shielding from ambient magnetic fields. The light is coupled to the detectors with an aluminum feedhorn block with a three step Potter\cite{Potter1963} style profile to provide a more symmetric beam in both polarization directions than previously achieved with conical feedhorns\cite{Rownd2003}. Each feedhorn pixel contains two orthogonally oriented MKIDs  (See Fig. \ref{fig:horn_layouts}). BLAST-TNG will serve as a pathfinder instrument for MKIDs \cite{Day2003}, which have never been flown before. The development of MKID arrays for astronomy is an extremely active area of detector research, and flight testing them will be a significant milestone.

Additional details of the instrument design including the electronics, pointing system, control system, and detector readout can be found in Ref. \citenum{Galitzki2014b}.

\input{HornArrays_v1}

\subsection{Cryogenic Performance}

The BLAST-TNG cryostat\footnote{Precision Cryogenic Systems Inc., 7804 Rockville Rd., Indianapolis, IN 46214} was delivered to the University of Pennsylvania (UPenn) in February 2015. Over the following year the thermometry, refrigerator systems, re-imaging optics, and the 250\mum\ detector array were integrated and tested in increments to allow us to characterize the performance of the cryogenic system and debug systems as they were added. The cross section of the cryostat can be seen in Figure \ref{fig:cryostat}. A thermal model was developed to describe the expected steady-state thermal load on each temperature stage. The thermal model took into account radiative loading as well as conduction through the G10 mechanical supports, cabling, motor axles, the helium fill tube, and the 1\,K refrigerator exhaust tube (See Table \ref{tab:therm_model}). 

During testing, the VCSs were observed to reach a steady temperature of $\sim$65\,K and $\sim$165\,K. The steady state load on the 4\,K cold plate was measured to be $\sim$340\,mW, as determined by the measured helium boil-off rate of the cryostat. The loading corresponds to a 22.5 day hold time for the 250 liter tank with an approximate boil-off rate of 11 liters of liquid helium per day. The observed loading is approximately 40\% larger than the predicted loading of 240\,mW from the thermal model used in the design of the cryostat. We believe our excess loading is due in part to un-modeled light leaks in the MLI blanket around fixtures and feedthroughs, as suggested by SPIDER\cite{Gudmundsson2015}, but also due to unexpected complexities in the thermal behavior of G10 material, which provides the mechanical support in the BLAST-TNG cryostat. 

\input{cryostat}


By adjusting the conductive model of the G10 we can account for most of the additional loading observed. This observation has prompted us to replace the G10 sections between the 165\,K and 65\,K stages and between the 65\,K and 4\,K stages with thinner walled material. The change will decrease the effective loading on the 4\,K stage to bring the performance in line with our target 28 day hold time. It should also be noted from Ref. \citenum{Gudmundsson2015} that the thermal loading at balloon flight altitudes is observed to be less due to the cooler temperature of the cryogenic vessel and the reduced optical loading through the window which can increase the hold times of cryostats during flight.

As a baseline for comparing cryostat performance a figure of merit was developed in Ref. \citenum{Holmes2001} that divides the radiative loading, $H$, by the cryogen depletion rate, $R$. For the SPIDER cryostat along with many satellite cryostats, $H/R \approx 60$\,W\,Days/L whereas the BLAST-TNG cryostat ranks slightly higher with $H/R\approx 230$\,W\,Days/L using the 250\,L cryogenic volume, a hold time of 22.5 days, and the approximately 6\,m$^2$ surface area of the 4\,K shield.

\subsection{Detector Performance}

The 250\mum\ MKID detectors have been shown to be photon noise limited in a 7 pixel test array that was feedhorn coupled to a variable blackbody source. Additional details of the tests and results can be found in Ref. \citenum{Hubmayr2015}. The noise equivalent power (NEP) of the detectors was fit using a three component noise model that included recombination noise and the photon noise on top of a flat background noise level. The fit determined that the detectors were limited by photon noise in the range 1\,pW to 20\,pW, which comfortably encompasses the expected BLAST-TNG flight loading on the detectors of $\sim$5 to $\sim$14\,pW. The fit to the noise model also produces an estimate of the optical coupling efficiency which was determined to be $\sim 70\%$. 

The 250\mum\ (1.2\,THz) array is designed to have a bandpass from 325 to 175\mum\ (1.0 to 1.4\,THz) which is defined on the long wavelength end by the feedhorn profile and on the short wavelength end, by a low pass filter mounted in front of the feedhorn block. The bandpass was tested using a hot $1050^\circ$\,C thermal source coupled to a Fourier Transform Spectrometer (FTS) that filled the feedhorn beam.  The tests determined that the edges of the bandpass for both polarization directions in a single pixel to be in agreement with the design of the filters and feedhorns\cite{Dober2015}.

The MKID's polarization performance has also been tested and is described in detail in Ref. \citenum{Dober2015}. The tests were performed with a source chopped between $1050^\circ$\,C and $20^\circ$\,C with a wire grid polarizer mounted between the source and the detectors. The polarizing grid is then rotated to produce a sinusoidal response at the detectors with orthogonal detectors 90 degrees out of phase. Fits to these measurements determined the cross-polar signal for both detector directions in a pixel to be $<$3\%, demonstrating the effectiveness of feedhorn coupled MKIDs for submillimeter polarimetry measurements.

Additional details of the MKID design and performance can be found in Ref. \citenum{McKenney2016}.

\subsection{Readout Electronics}

The BLAST-TNG readout is a highly multiplexed 1024 channel digital spectrometer which is the first of its kind to have been developed for the second generation Reconfigurable Open Architecture Computing Hardware (ROACH-2)\cite{Werthimer2011}. The ROACH-2 architecture is based on a field programmable gate array (FPGA), and is a product of the Collaboration for Astronomy Signal Processing and Electronics Research (CASPER). The ROACH is used in conjunction with a DAC/ADC board that was designed for ARCONS\cite{McHugh2012}. Building on the legacy of previous KID readouts created for the ROACH platform, the BLAST-TNG firmware performs coarse and fine channelization of 512\,MHz of RF bandwidth. The resulting channels are $\sim$100\,kHz wide and can be read out at rates of 200\,Hz to 500\,kHz. BLAST-TNG uses five ROACH-2 boards, three for the 250\mum\ array and one each for the 350 and 500\mum\ arrays, along with a suite of IF components and single board computers. The readout hardware is housed in the `ROACH-2 Motel', a custom enclosure designed for BLAST-TNG. Control software is written in Python, and will be ported to C for integration into the flight software. Details of the system’s overall performance are forthcoming. 
 

\subsection{Polarimetry} 
\label{sec:polarimetry} 

\input{hwpr_v1}

The primary scanning strategy during flight will be a slow raster scan, which works well for extended sources. A typical BLAST-TNG raster will scan across targets in azimuth at a speed of approximately $ 0.5^\circ{\rm s}^{-1}$ with an elevation scan speed calculated to change the elevation by 1/3 the array FOV in one crossing of the target in azimuth. A Half Wave Plate (HWP) is used to modulate the polarization signal, so that each pixel samples {\it I}, {\it Q}, and {\it U} multiple times during the course of a scan which allows us to control for polarization systematics. The HWP is stepped between four set angles (0$^\circ$, 22.5$^\circ$, 45$^\circ$, and 67.5$^\circ$) after each completed scan of a source in elevation. The HWP used in BLASTPol\cite{Moncelsi2014} had a 10\,cm diameter aperture and was made from five layers of 500\mum\ thick sapphire with an anti-reflective coating. However, the technology did not scale easily to larger apertures necessitating a new approach. The HWP developed for BLAST-TNG uses metal mesh filter technology\cite{Ade2006} to make a large diameter, $\sim$18\,cm, HWP (See Fig. \ref{fig:hwpr}). These types of HWP have demonstrated broad-band transmission at THz frequencies\cite{Pisano2012, Pisano2014}.

The detector rows are oriented parallel to the nominal scan direction with the MKID orientation rotated by 45 degrees in adjacent pixels along the scan direction. This alignment allows for sampling of both {\it Q} and {\it U} Stokes parameter on a timescale that is shorter than the array's common-mode $1/f$ noise, creating a redundant polarization modulation feature. The sampling timescale of the Stokes parameter is $\sim$0.025 s, which is determined by the detector separation, $\sim$45\arcsec\ at 250\mum, and typical scan speed, $\sim 0.5^\circ{\rm s}^{-1}$.

\section{Conclusion} 
\label{sec:conclusion} 

BLASTPol successfully demonstrated the viability of balloon-borne telescopes to explore the polarization of the submillimeter sky and has led directly to the development of BLAST-TNG.  Progress in the construction of the new instrument is on schedule and is proceeding rapidly. The cryogenics system has been thoroughly tested and is in a near flight ready status. Additionally, the detector technology is well understood and has undergone extensive testing at NIST prior to delivery to UPenn. The 250\mum\ array has been installed in the cryostat and has been successfully incorporated into the readout architecture with observed optical sensitivity during cryogenic operations. We expect to demonstrate the detector's target sensitivity in the flight cryostat during the summer of 2016. The 350 and 500\mum\ arrays and feedhorn blocks are being fabricated for testing and installation by fall 2016. The new BLAST-TNG instrument will give us magnetic field morphology maps at unprecedented resolution that span entire molecular cloud structures, allowing us to link the full-sky polarimetry maps of {\it Planck} with the high-resolution, small-area polarimetry maps, of telescopes such as ALMA. This will provide strong constraints on models of magnetic fields and turbulent interaction within magnetic fields as well as increasing our understanding of dust grain models. 

\section{Acknowledgments} 

BLAST is funded by NASA through grant number NNX13AE50G S08. Detector development is supported in part by NASA through NNH13ZDA001N-APRA. Brad Dober was funded by a NASA Earth and Space Science Fellowship NNX12AL58H S02. The BLAST-TNG collaboration would like to acknowledge the Xilinx University Program for their generous donation of five Virtex-6 FPGAs for use in our ROACH-2 readout electronics.
Peter Ashton was supported through Reach for the Stars, a GK-12 program supported by the National Science Foundation under grant DGE-0948017.
We would also like to thank the Columbia Scientic Balloon Facility (CSBF) staff for their continued outstanding work.

\bibliography{BLASTTNG_SPIE2016} 
\bibliographystyle{spiebib} 

\end{document}

%% file: blast-tng_model.tex
\begin{figure}[t]
  \centerline{
    \includegraphics[height=4in]{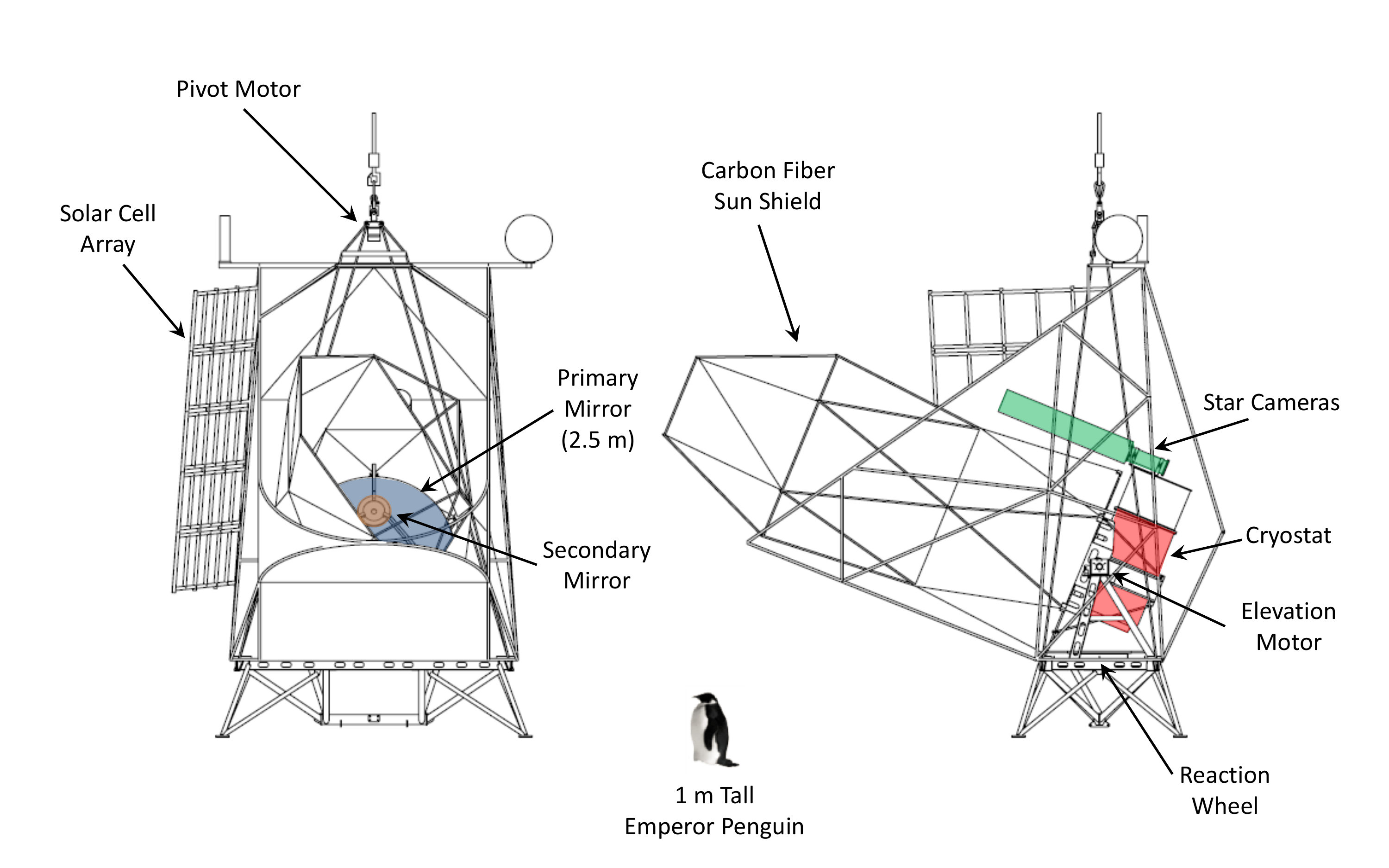}
  }
  \caption[Schematic view of the BLAST-TNG Telescope Configuration]{The front and side views of the BLAST-TNG telescope in its flight configuration. The cryostat, mirror optics bench, and star cameras are attached to an inner frame that moves in elevation. An extensive carbon fiber Sun shield also attaches to the inner frame to shield the optics at our closest pointing angle of 35$^\circ$ to the sun. \label{fig:model}}
\end{figure}

%% file: fgr_optic_v1.tex
\begin{figure}[t]
  \centerline{
    \includegraphics[height=2.5in]{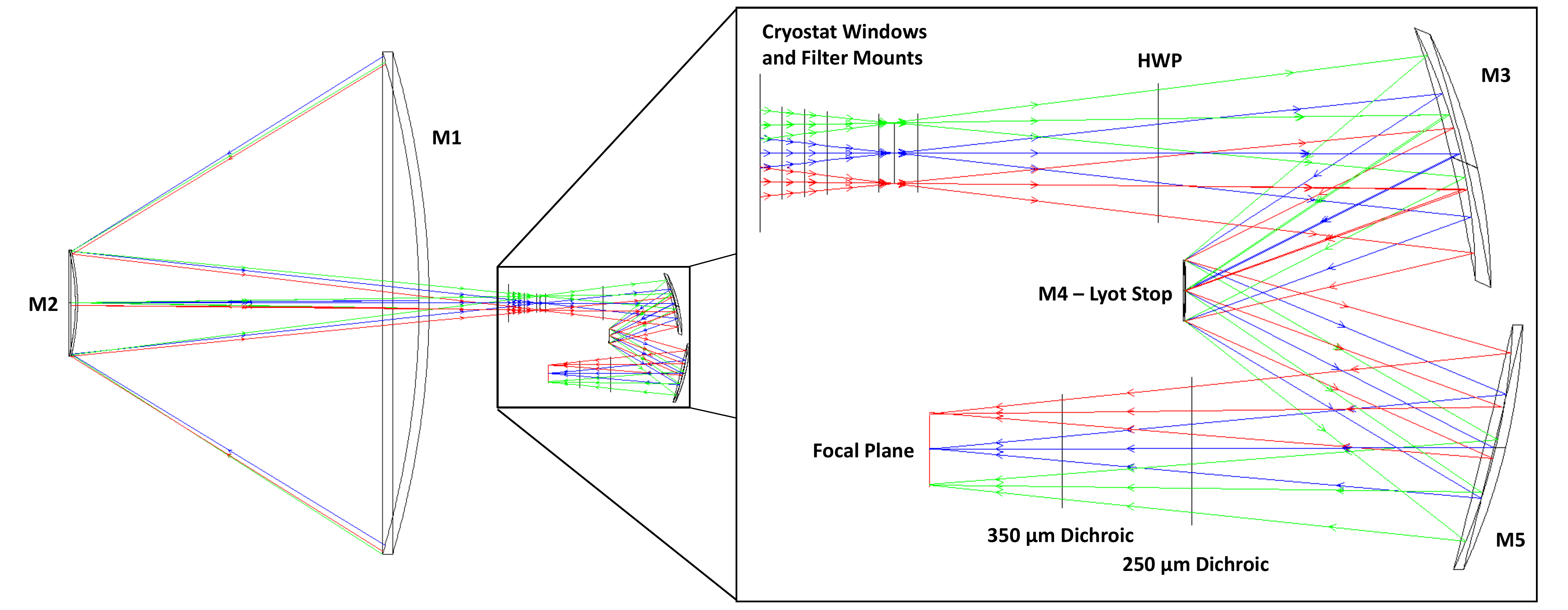}
  }
  \caption[BLAST-TNG optics design]{\label{fig:optics} Side view of the optical configuration of BLAST-TNG with a detailed view of the 4\,K cold optics. The telescope is an on axis Cassegrain that feeds into a modified Offner relay. M3, M4, and M5 are spherical mirrors with M4 acting as the Lyot stop for the telescope with a blackened hole that shadows the secondary mirror. Within the hole at the center of M4 is a calibrator lamp that provides an absolute calibration during flight operations to monitor responsivity drifts in the detectors. There are two dichroics that split the beam to the 250 and 350\mum\ arrays. The 250\mum\ dichroic is tilted at 22.5$^\circ$ to the optical axis while the 350\mum\ dichroic is tilted at 30$^\circ$ to the optical axis. Only one of the three focal planes is shown. The half wave plate is inserted between the Cassegrain focus and M3.}

\vspace{0.25cm}

\input{tab_optic_v2}	

\end{figure}

%% file: tab_optic_v2.tex

\newcommand\T{\rule{0pt}{3ex}}       
\newcommand\B{\rule[-2ex]{0pt}{0pt}} 
\vspace{3 mm}
\begin{center}
\caption{Summary of BLAST-TNG Optics Characteristics\label{tab:optica}}
\vspace{3 mm}
\renewcommand{\arraystretch}{1.2}%
\begin{tabular}{lccccc}
\hline
\hline
\T Geometrical Charac. & M1 & M2 & M3 & M4 & M5 \B \\
\hline
\T Nominal Shape & Paraboloid & Hyperboloid & Sphere & Sphere & Sphere\\                                                       
Conic Constant & $-$1.0 & $-$2.38 & 0.000 & 0.000 & 0.000\\
Radius of Curvature & 4.132$\,$m & 1.210$\,$m & 655.6$\,$mm & 376.5$\,$mm & 749.4$\,$mm\\
Aperture & $\varnothing$2.5$\,$m & $\varnothing$0.573$\,$m & $\varnothing$28$\,$cm & $\varnothing$7$\,$cm & $\varnothing$28$\,$cm \B \\
\hline
\end{tabular}
\end{center}

%% file: HornArrays_v1.tex
\begin{figure}[t]
  \centerline{
    \includegraphics[width=7in]{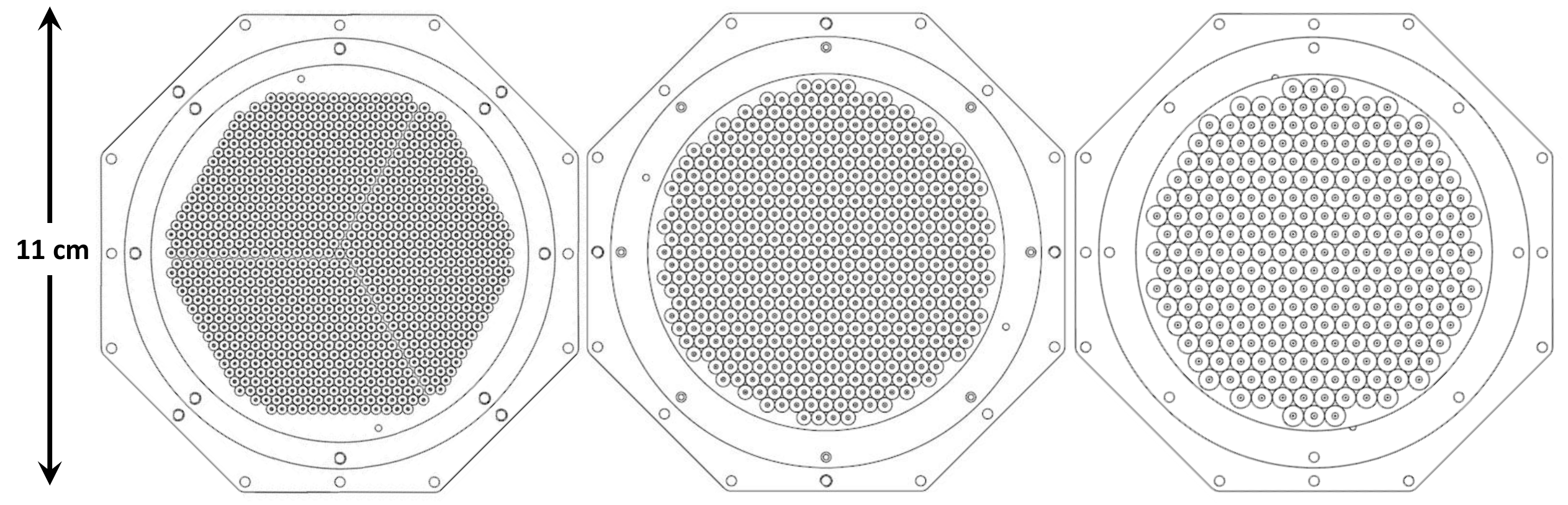}
  }
  \caption[The three horn array layouts]{\label{fig:horn_layouts} A top image of the feedhorn block designs corresponding to the 250, 350, 500\mum\ arrays (from left to right) with 759, 475, 230 pixels, respectively. Each pixel contains two orthogonally oriented MKIDs.}
\end{figure}

%% file: cryostat.tex
\begin{figure}[t]
  \centerline{
    \includegraphics[height=3.5in]{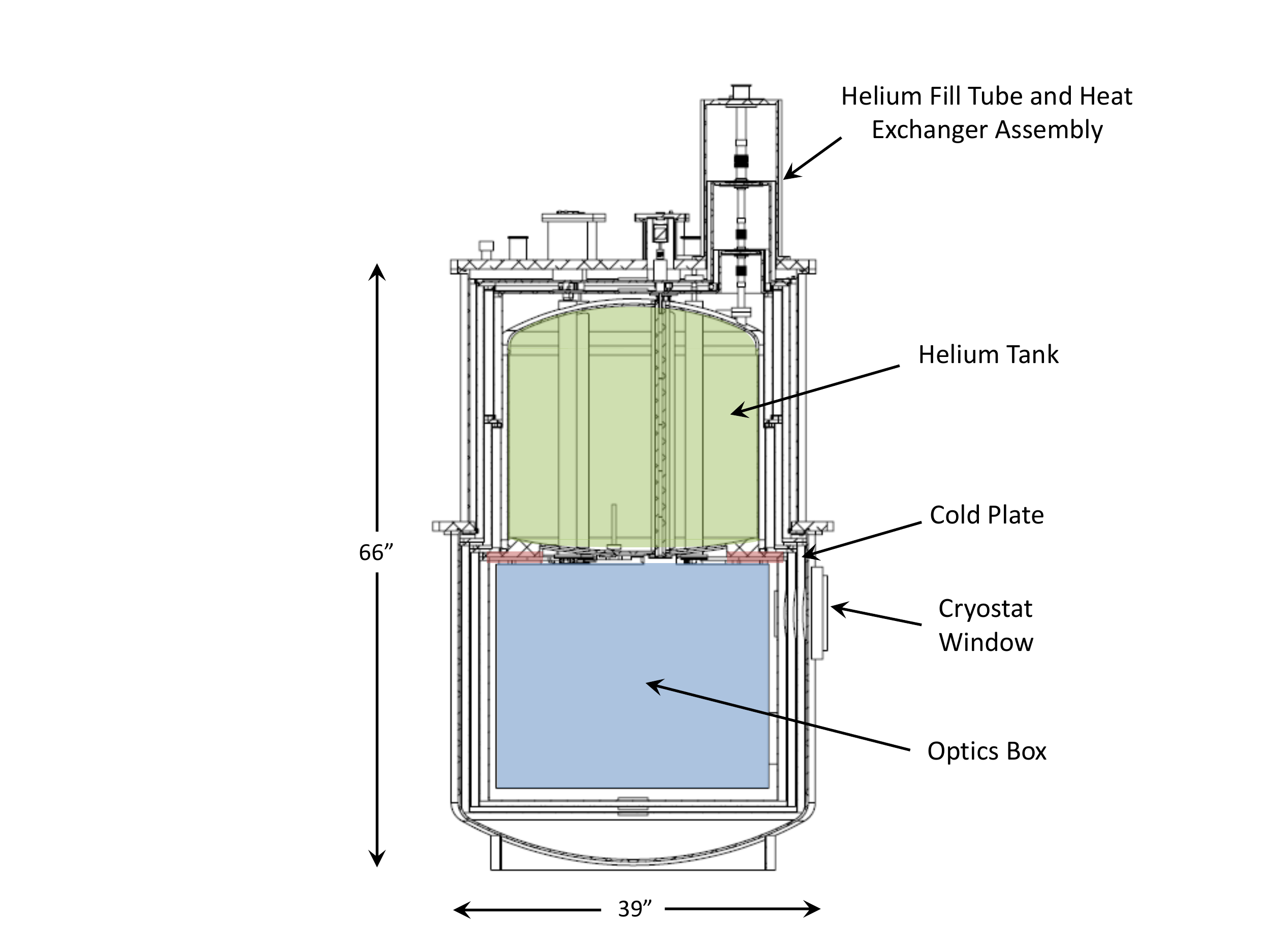}
  }
  \caption[BLAST-TNG cryostat cross-section]{\label{fig:cryostat} Cross section view of the BLAST-TNG cryostat. The liquid helium cryostat includes two helium vapor heat exchangers that cool two shields to provide thermal isolation of the cold optics. The inner vapor cooled shield is kept at $\sim$65\,K and the outer one at $\sim$165\,K. The cryostat has a predicted hold time of 28 days.}
	
\vspace{0.25cm}

\input{tab_therm_model_v1}

\end{figure}

%% file: tab_therm_model_v1.tex

\vspace{3 mm}
\begin{center}
\caption{List of the percentages of primary contributions to the thermal load at each of the main stages \label{tab:therm_model}}
\vspace{3 mm}
\renewcommand{\arraystretch}{1.2}%
\begin{tabular}{lccccc}
\hline
\hline
\rule{0pt}{3ex} &Radiative & Conductive & Plumbing & Cables & Misc. \rule[-2ex]{0pt}{0pt} \\
\hline
VCS2 Stage (165\,K) & 77\% & 16\% & 6\% & 1\% & $<$1\% \\
VCS1 Stage (65\,K) & 22\% & 50\% & 22\% & 6\% & $<$1\% \\
LHe Stage (4\,K) & 24\% & 50\% & 11\% & $<$1\% & 14\% \\
&&&&&\rule[-2ex]{0pt}{0pt} \\                                                    
\hline
\end{tabular}
\end{center}

%% file: hwpr_v1.tex
\begin{figure}[t]
  \centerline{
    \includegraphics[width=5in]{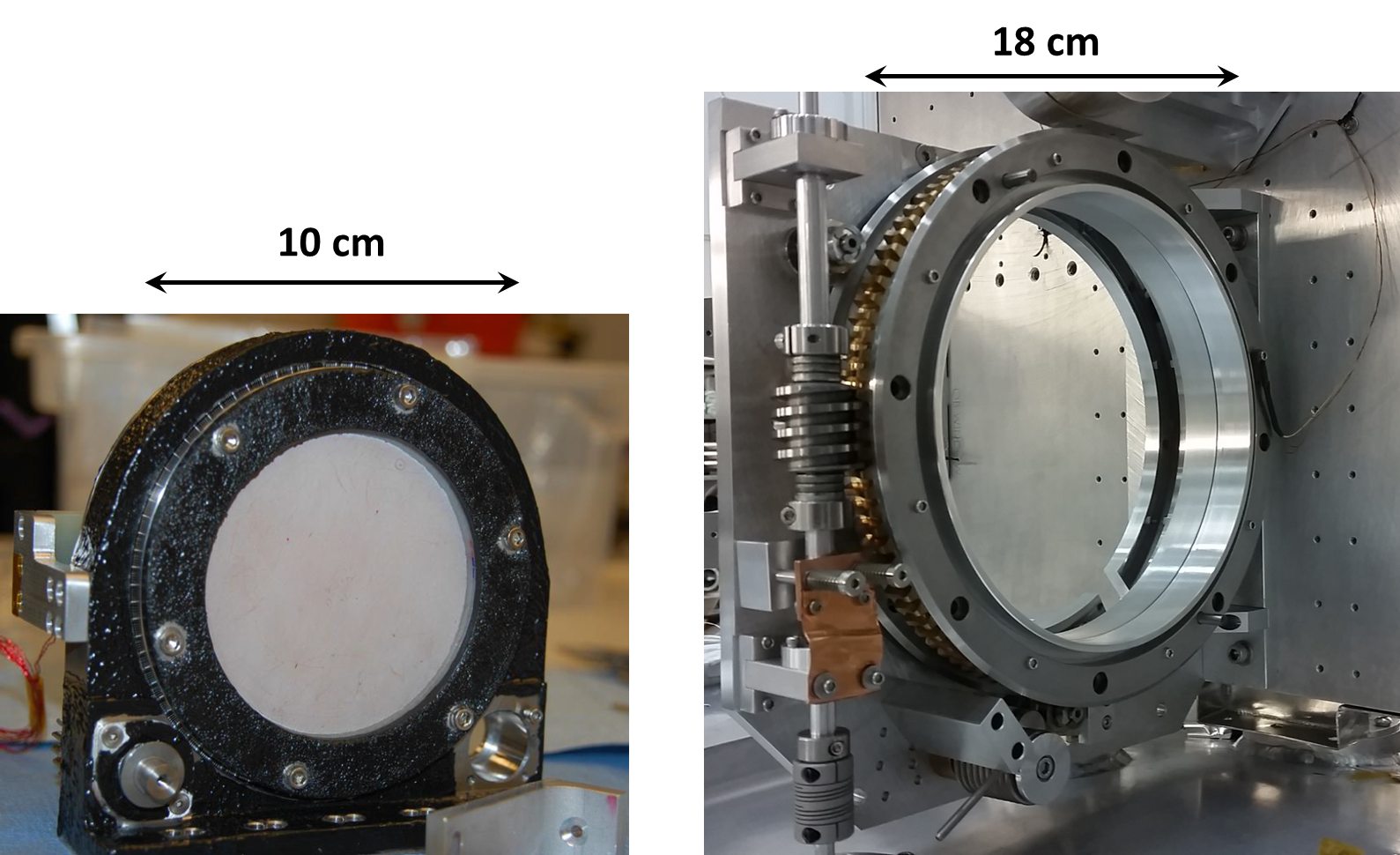}
  }
  \caption[Half-wave plate design comparison]{\label{fig:hwpr} Left: An image of the half wave plate and rotator mechanism that was used in BLASTPol\cite{Moncelsi2014}. Right: An image of the new HWP rotator mechanism with a larger clear aperture of $\sim$18\,cm. The rotator is bolted to the optics bench which is in turn mounted on the cold plate of the helium tank. Rotation of the HWP is driven by a worm gear that is coupled to a motor mounted to the 300\,K lid of the cryostat with a G10 shaft to provide thermal isolation.}
\end{figure}